# Experiments reveal extreme water generation during planet formation

Miozzi F.[1*], Shahar A.[1], Young E.D.[2], Wang J.[1], Steele A.[1], Borensztajn S.[3], Vitale S.M.[1], Bullock E.S.[1], Wehr N.[3], Badro J.[3]

[1] Earth and Planets Laboratory, Carnegie Institution for Science, Washington D.C., 20015, USA.
[2] Department of Earth, Planetary, and Space Sciences, University of California Los Angeles, Los Angeles, CA, USA.
[3] Université Paris Cité, Institut de Physique du Globe de Paris, CNRS, Paris, France.

*fmiozzi@carnegiescience.edu

Abstract
The most abundant type of planet discovered in the Galaxy has no analogue in our Solar System and is believed to consist of a rocky interior with an overlying thick $H_2$ dominated envelope. Models have predicted that the reaction between the atmospheric hydrogen and the underlying magma ocean can lead to the production of significant amounts of water. The models suffer however from the current lack of experimental data on the reaction between hydrogen and silicate melt at high pressures and temperatures. Here we present novel experimental results designed to investigate this interaction. Laser heating diamond anvil cell experiments were conducted between 16 and 60 GPa at temperatures above 4000 K. We find that copious amounts of hydrogen dissolve into the silicate melt with a large dependence on temperature rather than pressure. We also find that the reduction of iron oxide leads to the production of significant amounts of water along with the formation of iron-enriched blebs. Altogether, the results predict that the typical processes attending planet formation will result in significant water production with repercussions for the chemistry and structure of the planetary interior as well as the atmosphere.

Introduction
Hydrogen, the principal constituent of primary atmospheres, plays a fundamental role in planetary accretion. Its interaction with the planetary interior—particularly the hot magma ocean—establishes the boundary conditions for chemical evolution, thereby influencing the planet's structure, dynamics, and marking a crucial stage in its formation [1,2]. Notably, two key reactions are proposed to dominate this stage of accretion. The first involves the reduction of FeO in the melt through reaction with atmospheric $H_2$, to form Fe and $H_2O$ according to FeO + $H_2$ = Fe + $H_2O$ [2,3]. The second entails the ingassing of hydrogen into the silicate melt via the equilibrium: $H_2^{gas}$ = $H_2^{melt}$ e.g., [4,5]. The extent of change that such reactions can cause on the interior depends mainly on hydrogen's solubility in the melt and how it changes with pressure and temperature. Constraining solubility is fundamental as the silicate melt can act as a sink for hydrogen, trapping it in the mantle. There, among other reactions, hydrogen can reduce other species to produce water, which can then partition between the silicate melt and the overlying atmosphere. Indigenous formation of water by this reaction has been recognised as a possibility [6,7] but is yet to be established in the laboratory.

Recent modelling studies have investigated the equilibration of a magma ocean with a hydrogen-rich atmosphere, focusing their application to Sub-Neptunes, the most abundant type of exoplanet discovered and with no analogue in the Solar System [3,7,8]. Many Sub-Neptunes are thought to have molten rocky interiors enveloped by dense $H_2$ atmospheres. These atmospheres act as a thermal blanket, keeping magma oceans hot for billions of years[9,10] during which interaction between these two reservoirs can lead to water formation[8,11]. The bulk of the experimental work relevant to this system, to date has focused on: i) assessing solubility in evolved melt compositions [4,12], at low pressure and often without iron; ii) the reaction of simplified systems with hydrogen as MgO + Fe [13] or $Fe_2O_3$ and $(Mg_xFe_x)O$[14]; or iii) investigating hydrogen partitioning into the metal core of terrestrial planets [15,16]. Experimental challenges arise from the extremely high temperature required to melt silicates with a primordial composition in the required pressure range, working with hydrogen gas without breaking diamonds or losing the sample and recovering intact samples with sufficient area to perform analytical analyses.

Here we present novel experiments investigating the interaction between a hydrogen-bearing atmosphere and a primitive iron-bearing magma ocean at high pressure and high temperature. The silicate was equilibrated with hydrogen gas and heated to temperatures above melting. The experiments were designed to investigate if there is a combined effect of FeO reduction and H$_2$ ingassing, as hypothesized in previous studies [17,18], and assess hydrogen solubility in the silicate melt. Our results represent the first experimental evidence of hydrogen ingassing and FeO reduction from a primitive mineralogical system from the high-pressure interaction with H$_2$ and the formation of an Fe-dominated phase. We report a net increase in the hydrogen content of the melt with respect to previous measurements at lower pressure and temperature [12]. Finally, we show evidence of the presence of a free fluid phase in the recovered experiments. Taken together the results provide the first insights into the processes taking place during atmosphere-magma ocean interaction in the early stages of planetary evolution and in the interior of rocky planets.

Chemistry and hydrogen content
Laser heated diamond anvil cell experiments were performed at pressures between 16 and 60 GPa. Platelets of pyrolytic composition were loaded in an Ar-H$_2$ mixture, heated at temperatures above 4000 K [19,20] (Tables 1 and S1) and recovered and prepared for NanoSIMS analyses with a newly developed protocol (Methods and Supplementary information).
In the three samples recovered from lower pressure experiments (i.e. Exp 11-13) the melt pocket crystallized in concentric layers around a resin-filled cavity. Overall, the silicates show minimal fracturing, limited to the boundaries between layers, and remain well consolidated (e.g., Exp_12 in Figure 1). The crystallized solids assemblage occupies the outermost layer, while the residual melt occupies the innermost layer (Text S3 and Table S2) and is in contact with the cavity. At the boundary, rounded Fe-enriched blebs are observed (Text S4, Figure S3, S4, S5). The cavity has an oblate shape in most samples (Figure S5). The observation of rounded central cavities is novel for diamond anvil cell experiments as usually the melt or a metallic phase occupies the central portion of the sample [e.g., 21,22]. At higher pressure, in Exp_09 the silicates present more fractures. The cavity is flat and elongated (Figure S6). The Fe-enriched blebs are almost dome-shaped and the border with the silicates is irregular with respect to the lower pressure samples (Figure S4). Finally, the highest-pressure sample (Exp_10) displays extremely different characteristics. The silicates present an extensive network of fractures and are poorly consolidated (Figure S6). The cavity is much smaller and little, irregular, and randomly distributed cavities are also visible. The Fe-enriched bleb has an irregular shape and is intertwined with the silicates.
In all of the samples homogeneously distributed nanometric bubbles are visible in the melt. The melt also displays a higher Ca content with respect to the crystallized solids

Table 1: Experimental conditions and results of NanoSIMS analyses

| Sample | Pressure before heating (GPa) | Temperature (K) ± 150 | Measured water (wt%) | Combined uncertainty |
|---|---|---|---|---|
| Exp_09 | 46.2 | 4500-4600 | 5.3 | 0.1 |
| Exp_10 | 60 | 4700 | 5.4 | 0.3 |
| Exp_11 | 16 | 4000 | 5.6 | 0.1 |
| Exp_12 | 22.4 | 4100 | 6.5 | 0.1 |
| Exp_13 | 30.7 | 4100-4200 | 5.9 | 0.1 |
| Exp_10s (crystallized solids) | 60 | 4700 | 0.81 | 0.01 |

(see Figure S2 for an example). Importantly, FeO is absent from the melt in all samples excluding Exp_13 (Figure S3). All the Fe-enriched blebs show some texturing (Text S4 and Figure S4).
The hydrogen content was measured with a Cameca NanoSIMS 50L (Table 1 and Figure 1 and S7). The instrument detects dissolved hydrogen species that are then reported as water (water equivalent), regardless of the speciation that H might have in the sample (i.e., H$_2$, H$_2$O or OH$^-$). Accordingly, we provide NanoSIMS results as water and possible alternatives to water equivalent are reported in Table S2. In all of the samples, between 5 and 6 wt% water was measured in the quenched silicate melt (Table 1 and Figure S8) and, 1 wt% water was measured in the crystallized solids of sample Exp_10 (Figure S9).

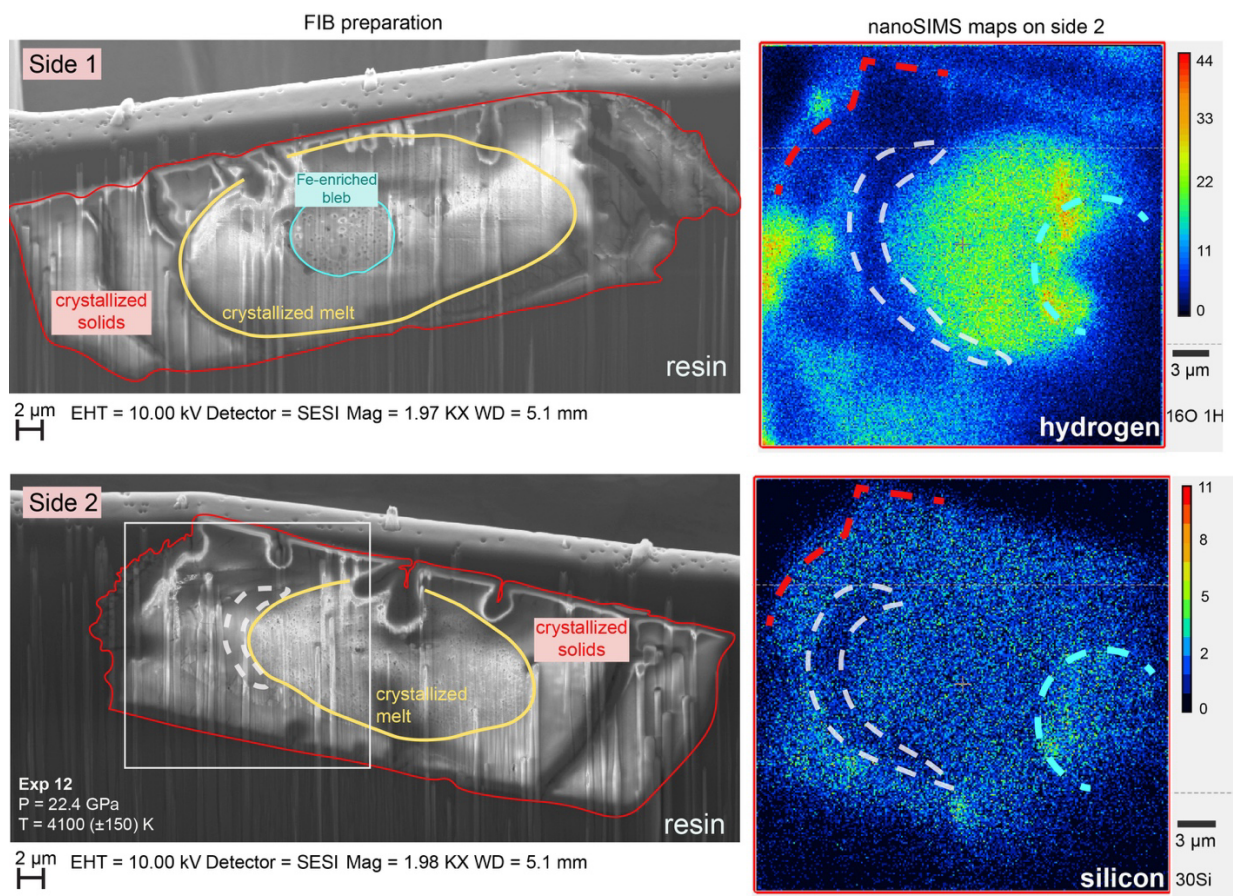

Figure 1: Left side: Scanning Electron Microscope (SEM) images collected on the two sides of the sample during milling with the Focused Ion Beam (FIB). Red lines define the sample. The crystallized melt is characterized by extensive presence of nanometric bubbles. The image of the sample's section with the visible cavity is in Figure S5.
Right side: Maps of hydrogen and silicon collected with the NanoSIMS on a portion of the sample (shown as a white box in the SEM image). Dashed lines serve as a guide for the eye. Red shows the sample boundary, white shows a portion of the crystallized solids and blue shows the portion behind a Fe-enriched bleb. The color bar represents the intensity of the signal from the corresponding mass for each pixel. In the hydrogen map the sample shows extremely high counts in the melt.

Discussion

The measured water and presence of nanometric bubbles, along with the absence of FeO, and formation of Fe-enriched blebs confirm hydrogen ingassing in the melt and FeO reduction with water production. Accordingly, two processes occur during silicate melt-hydrogen interaction. One is hydrogen incorporation in the melt, as evidenced by the NanoSIMS data, where the dissolved $H_2$ can be in the molecular form or as hydroxyl groups. The second is oxide reduction, particularly FeO, and water production, forming a free fluid phase. The latter is heavily supported by the observation of empty cavities in the lower pressure samples. At the experimental conditions the water phase would coexist with the melt and the Fe-enriched blebs as separate phases. It would then escape the sample during recovery at ambient conditions, leaving behind an empty cavity . If all the FeO is reduced, ~6wt% FeO in the starting material would yield ~1.6 wt% water. This represents a lower bound, as other oxides might also reduce and contribute water. NanoSIMS hydrogen measurements similarly reflect a lower solubility limit. The empty bubbles in the silicate melt, likely a product of temperature quench [13], contained exsolved volatiles that were permanently lost upon FIB milling.

We interpret our observations of Exp_10 (Figure S6) as quench from a single-phase, resulting in a less consolidated sample with a wide-spread network of fractures and irregular boundaries between the different phases. We infer that at this higher conditions the system likely encountered a solvus (binodal), the boundary above which all phases become miscible and exist as a single phase [23]. Such interpretation is consistent with previous observations on simplified systems [13]. In this framework the P-T conditions for Exp_09 should be close to complete miscibility as it shows features of both the low- and high- pressure samples.

To understand the effects of planetary sizes and conditions on hydrogen solubility into a silicate melt, we parameterize the dependence from pressure and temperature. In the absence of direct evidence of hydrogen's speciation from our experiments, we model it as $H_2$ following previous studies [4,12]. We combined our data with the low pressure solubility data for basaltic melt [12] and considered only $H_2$ diffusion in the melt, equating the chemical potentials for $H_{2,gas}$ and $H_{2,melt}$ (see Methods for details). Using $P$ and $T$, $lnX_{H_2 melt}$ from the experiments, multivariate regression over all the experiments results in $\Delta G^{0,*}$ = 30.577 +- 6.22 kJ/mol, $\Delta V$ = 1.93e$^{-7}$ m$^3$/mol and $fH_2$ = 40979 Pa, the latter corresponding to a mole fraction of the hydrogen in the gas of 1.86e$^{-6}$, suggesting the hydrogen in the gas was almost fully dissolved into the sample.

Hydrogen solubility varies significantly with temperature but only weakly with pressure (Figure 2), showing a slight decrease in the mole fraction of $H_2$ in the melt with increasing pressure at constant temperature. The solid-melt partitioning coefficient $D_{sil-melt}$ = 0.151 (± 0.186) from Exp_10, shows that hydrogen preferentially enters the melt during crystallization.

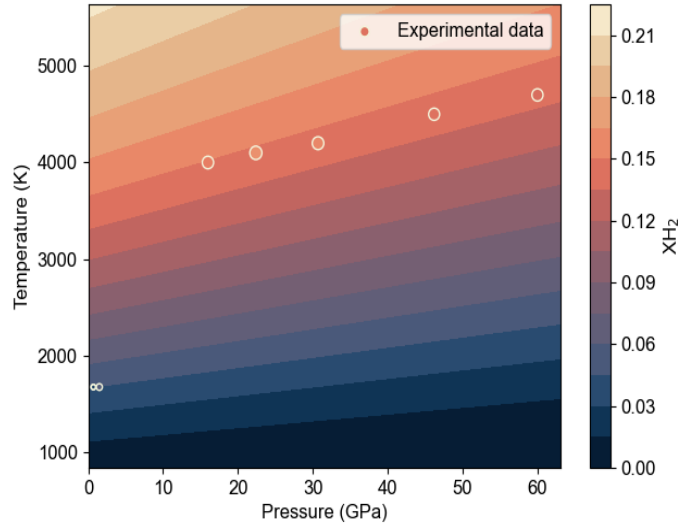

Figure 2: The dependance of hydrogen solubility on pressure and temperature. Colored contours are the mole fraction of $H_2$ in the melt recalculated with the parameters fitting the equations. Symbols represent the content measured in the experiments (see also Figure S8). The data below 10 GPa are from Hirschmann et al. (2012)[12]. Symbols' size is scaled with hydrogen content.

Implications for planetary interiors
Our results advance our understanding of water production during formation and evolution of planetary bodies and are key to elucidating aspects of Sub-Neptunes interiors and atmospheres. Previous experimental works focused on a smaller pressure range, investigating only simplified systems in a qualitative manner. Theoretical work, on the other hand, investigated the interaction in the complex system, but key parameters such as the solubility of hydrogen at relevant conditions were still unknown. Here we provide an experimentally derived framework for the interaction of a multi-oxide mineralogical system with hydrogen. We show conclusively that a significant amount of hydrogen is dissolved in the melt, that FeO is reduced to Fe, and that the water produced from the reaction forms a free fluid phase at lower pressure and becomes miscible with the silicates above 45 GPa (Figure 1, S5, S6, S9). Additionally, reduction of other oxides present in the melt can represent an additional pathway for water formation [8,17]. Qualitatively our experimental results are aligned with recent literature on exoplanets, predicting a high volume of endogenous water resulting from magma ocean-atmosphere interaction [11,24]. In planetary interiors, the amount of water produced, and its fate depends on the available oxides and crucially on the degree of interaction between the silicate melt and the hydrogen. For planetary bodies not massive enough to retain an atmosphere in the early magma ocean stage hydrogen ingress is likely limited [8]. However, once planets grow to on the order of about 0.25 Earth masses or greater [2], $H_2$-rich envelopes can be retained with magma oceans beneath, leading to the ingassing of substantial amounts of hydrogen and the production of endogenous water. Temperature will control hydrogen solubility, and the dynamic nature of the magma ocean will control oxide delivery to the surficial portions and reduction. It is possible to envision scenarios in which water gets released into the atmosphere creating a steam atmosphere or scenarios in which it is entrained in the magma ocean and accumulates at depth (Figure 3). Characterizing the dynamic magma ocean and the consequences for the budgets of volatiles is challenging. Nevertheless, our results underscore the importance of such investigations, as hydrogen and water might be significantly more prevalent than previously anticipated for planets accreting with a hydrogen atmosphere overlying molten interiors.

Understanding volatile-rich systems is pivotal to exploring a potential feature of Sub-Neptunes: complete miscibility between phases, forming one single mixed phase in the supercritical state in the mantle [18] (Figure 3). The boundary between a "standard" interior with separate phases and one in the supercritical state is the position of the solvus for the system in equilibrium with volatiles. Our results suggest complete miscibility, and thus conditions above the pyrolite + $H_2$ solvus at pressures above 45 GPa and temperatures above 4500 K for the hydrogen concentration of our experiments. Ab initio calculations performed on the $MgSiO_3$ + $H_2$ system report full miscibility in a pressure and temperature range consistent with our experiments [25].

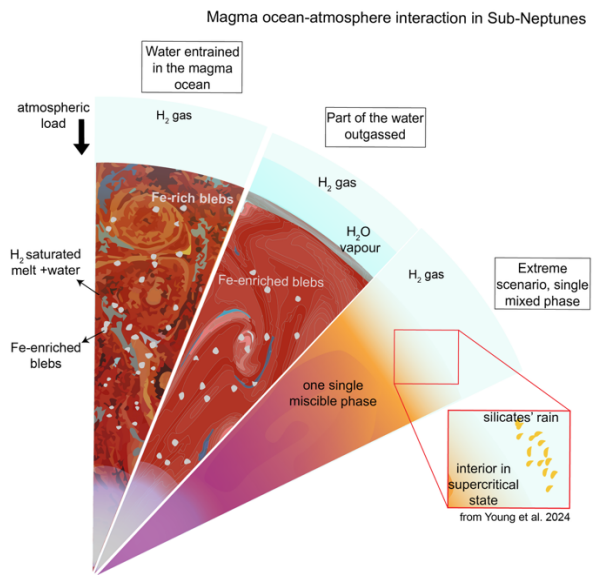

Figure 3: Left, the free fluids phase and Fe-enriched blebs are entrained in the magma ocean. Center: water is outgassed to create a steam atmosphere while the Fe-enriched blebs sink toward the center. Right: extreme scenario in which all phases are miscible.

Ultimately, along with water production, FeO reduction also implies the formation of an Fe-rich phase. Whether this phase sinks to the core depends on its chemical and physical properties as well as the planet's mass. For smaller planets the iron-rich phase is expected to sink to the core (e.g., Earth). For more massive planets, where temperatures are higher the concentration of light elements in the Fe-rich phase formed through this process may render it so under-dense as to be neutrally buoyant[18].

Notwithstanding the uncertainties regarding complete miscibility, our experimental results reveal an alternative pathway for the incorporation of water into planetary interiors. The hydrogen solubility measured in the quenched silicate melt, shows that significant amounts of hydrogen can be stored in a magma ocean while water is being formed concurrently. The storage mechanisms may have unexplored yet potentially important effects on the physical and chemical properties of such systems and consequently on a planet's large-scale properties and habitability.

Methods
Diamond anvil cells (DAC) with culet size ranging from 300 to 200 μm were used for the experiments. A 200 μm thick rhenium foil was indented to a thickness between 30 – 45 μm and drilled with a laser drill system available at Carnegie's Earth and Planets Laboratory (EPL) [26]. Samples were produced with a gas-mixing aerodynamic levitation furnace [27] operating at 2173 K and with an Ar flow (PYR03) or a Ar-$H_2$ flow (92%Ar 8%$H_2$) for PYR05. The mm-sized glass beads were then sliced with a wire saw and double polished to reach a thickness of 25-40 μm. Chemical composition was measured with a JEOL JXA-8530F field emission electron microprobe analyzer, working with a 15 kV beam and 20nA current.
The platelets were gently cracked to produce samples with the desired dimensions and were loaded in the experimental chamber along with ruby spheres, used as pressure calibrants. In some cases, small KCl flakes were used to prevent the sample from sticking to the diamond. All the cells were gas loaded at EPL, with an Ar-$H_2$ mixture (75%Ar, 25%$H_2$).

Laser heating experiments were conducted at the Institut de Physique du Globe de Paris (IPGP), France. The samples were heated from both sides using two laser branches that are linearly polarized and where each beam goes through a half-wave plate and a polarizer. The polarizers allowed us to independently change the power reaching each side of the sample, adjusting both sides to have the same temperatures.
Temperature was measured by spectroradiometry on a central 10 μm area. We fit both the Plank and Wien functions, and temperature is constantly monitored from both sides as shown in [20]. Samples Exp_09-13 were recovered by pouring epoxy resin (EpoTeck 301, from Epoxy Technologies) into the indentation of the gasket while the latter was still mounted on half of the cell. This was pivotal for our study as it allowed us to recover the sample while maintaining its structural integrity. This gives access to a full characterization of the structure and the chemistry of the entire sample, unlike previous studies involving water. After curing, part of the resin was removed with a needle to expose a sufficient portion of the culet at the bottom and cut the metal with the picosecond laser machining system available at IPGP. With this procedure the recovered samples are like standard DAC recovered samples (i.e., the Re-disk with the sample inside) except for the resin present on top of the experimental chamber. As having to go through resin is not ideal for FIB milling, the samples were mounted upside down on a carbon dot. Samples were gold coated and milled using the Auriga 40 Zeiss FEG $Ga^+$ FIB and FEG-SEM, equipped with a Bruker EDX detector at IPGP. The samples were brought to beam coincidence and trenches were made alternating the two sides until one of the two had a sufficiently big portion of melt without fractures. Quantitative chemical maps (calibrated EDX k-factors,[20]) were collected after every cut to monitor the evolution of the chemistry with the milling and identify when the central portion was reached. Before proceeding with the lift-out, Pt was deposited on top of the thick sections and choice were made in regard to where to cut laterally. In fact, the samples were too wide (~90μm) to be lifted entirely. Finally, Sections 40-60 μm width and 20-40 μm high were lifted and Pt-welded on a Cu-base TEM grid. The stage was then tilted to have the section perpendicular (within 12°) to the SEM column and elemental maps were collected for 30 minutes.

To recover the control sample, the cell was slightly opened to release the gasses from the experimental chamber and then placed into a canister with an inflatable membrane. Pressure was increased using ruby fluorescence to track the pressure evolution and the camera to optically monitor changes in the geometry of the experimental chamber. Above 12 GPa no changes were observed in the geometry of the experimental chamber but to be sure the cell was compressed to 25 GPa. The cell was decompressed at a slower rate to avoid cracking the sample.
Before milling the sample was analyzed with a WiTec confocal scanning Raman system (Alpha 300R) modified for DAC work. Both individual Raman spectra and images were collected on the sample within and outside of the DAC using a 488 nm laser excitation wavelength. Spectra were collected on a Peltier-cooled Andor EMCCD chip, after passing through a f/4 300mm focal length imaging spectrometer, typically using a 600 lines/mm grating. Spectra and maps were background subtracted, and cosmic ray removed before undergoing PCA analysis all using Witec Project 5 software.

The sample was recovered at EPL by cutting the central portion of the culet with the laser machining instrument. The disk was then mounted on a Si-wafer with silver epoxy. At EPL the sample was milled with a FEI Helios Plasma Focused Ion Beam working with $Xe^+$ following the same procedure as for the other samples milled at IPGP with two exceptions. First, a 10-15 µm tungsten layer was deposited on the entire disk before starting, to protect the front of the cut given the higher currents used with a $Xe^+$ FIB in respect to a $Ga^+$ FIB. Second, the higher working current and characteristics of the instrument made it possible to lift the sample with its full width, so instead of cutting trenches to reach the section of interest, the entire width of the disk was ablated. This procedure was not used for the samples milled at IPGP as ablating the rhenium portion of the disk with a $Ga^+$ FIB is not efficient. For NanoSIMS analyses the TEM grids with the thick sections were mounted on copper tape with the surface of interest looking upwards. Gas release from the glue component of the tape, in the NanoSIMS vacuum, is considered negligible (Nittler L., personal communication). As the instrument does not have access to speciation, we used Raman spectroscopy to try to determine the speciation of the hydrogen-bearing species in the melt but the analyses were not conclusive (Text S5).

NanoSIMS analyses:
NanoSIMS analyses were performed using a $Cs^+$ beam with 2nA current. For each analyzed portion a 15x15µm region was pre-sputtered to remove surface contaminants. The rastered area was then reduced to 10x10 µm with the area represented by the analyses being a 5x5 µm portion, due to the beam blanking used to reduce the effect of the crater's edge. Along with $^{16}OH^-$ and $^{30}Si^-$, specie $^{12}C^-$, $^{19}F^-$, $^{32}S^-$ and $^{35}Cl^-$ were also collected. The timeframe for data collection was 150s, however for one of the samples (Exp_09) we needed to cut the analyses at 70 second as a sudden drop in counts signaled that the sample was sputtered away.

We used synthetic forsterite and Suprasil $SiO_2$ glass to determine background and detection limit, and Alvin 519-4-1, 1833-1, 1846-12, 30-2, WOK28-3 basaltic glasses for water and other volatiles calibration. The $SiO_2$ measured in the standards was used to calibrate the acquired species/$^{30}Si$ ratios and define the calibration curve used to determine water content of the samples given the measured $SiO_2$ content (see also NanoSIMS data.xlsx). Hydrogen was calculated by dividing the counts obtained for water by a conversion factor calculated dividing two times the molecular weight of hydrogen against water's molecular weight. Uncertainties reported in Table 1 are the combined uncertainties that includes uncertainties of the collected data determined based on the counting statistics, backgrounds and repeated analyses of 519-4-1 standards, those on the measured $SiO_2$ content, as well as the one on the calibration curve. Careful consideration of the position of the analytical point was needed as for most of the samples the Fe-enriched blebs are observed on one side. Similarly, particular attention was given to selecting sections minimizing the number of fractures in the samples. This purpose was achieved in the lower pressure samples, it was harder for the high-pressure samples due to the lower volume and the important thinning of the sample along the compression axis.

Thermodynamic model
In using the data for basaltic melt from [12] it should be noted that the hydrogen contents at 1 and 3 GPa might be slightly different for pyrolite, as it has less network-forming cations (e.g., $Ca^{2+}$, $Al^{3+}$) and more network modifiers (e.g., $Mg^{2+}$) than basalt [12,28,29]. However, solubility is suggested to be less affected by composition for small molecular species [30], thus supporting the use of both datasets.

We parametrized the dependence of the solubility from pressure and temperature considering only the reaction of $H_2$ diffusion in the melt. Equating the chemical potentials for $H_{2,gas}$ and $H_{2,melt}$ leads to:

$$\mu_{H_2 melt}^{0,*} + RT \ln X_{H_2 melt} + \Delta \bar{V}_{H_2 melt} \cdot \Delta P = \mu_{H_2 gas}^{0,*} + RT \ln X_{H_2 gas} + RT \ln(P^*/P^{0,*})$$

where star denotes the pure $H_2$, 0 superscript denotes the standard state (i.e. 1 bar, 300 K), $\bar{V}_{H_2 melt}$ is the partial molar volume of $H_2$ in the melt, and we adopt pressure of the pure gas species, $P^*$, as a surrogate for fugacity of the pure species. Combining the standard-state terms for the pure species into the change in the standard-state molar Gibbs free energy $(\bar{G}^{0,*}_{H_2gas} - \bar{G}^{0,*}_{H_2melt})$ and replacing $ln\left(\frac{P^*}{P^{0,*}}\right) + lnX_{H_2gas}$ with $ln(\frac{fH_2}{P^{0,*}})$ yields:

$$lnX_{H_2melt} = -\frac{\Delta \bar{G}^{0,*}}{RT} - \frac{\Delta \bar{V}_{H_2melt}}{RT} \cdot \Delta P + ln\frac{fH_2}{P^{0,*}}$$

Using the hydrogen fugacity term to account for both the mole fraction of $H_2$ in the gas and the non-ideal-gas equation of state for the "gas" phase avoids assumptions about non-ideal behavior. We use this expression to derive typical values for the fugacities of hydrogen for the experiments as a group. This is justified by very low values obtained relative to experimental pressures, even when including $H_2$ fugacity coefficients that range from about $10^2$ to $10^4$ over our experimental conditions [31]. The $R^2$ for the fit is 0.91 based on experimental uncertainties. Comparison between the observed and modelled data is in Figure S10. The covariance matrix of the fit was used to calculate the 1 sigma errors reported for each parameter. The partial molar volume is smaller than the one reported by Hirschmann et al. (2012)[12] at 4 GPa and 3500 K (1.6 e$^{-5}$ m$^3$/mol and 7.3e$^{-6}$ m$^3$/mol respectively).

Acknowledgements

FM thanks T. Gooding for the discussions about epoxy resin and sample recovery procedures. G. Cascioli, E. Codillo, V. Dobrosavljevic, Z. Geballe and G. Criniti for the support and discussions.
This AEThER publication is funded by the Carnegie Institution for Science and the Alfred P. Sloan Foundation under grants G202114194 and G-2025-25284. The work and JB acknowledge funding from the European Research Council (ERC) under the European Union's Horizon 2020 research and innovation program (grant agreement no. 101019965— ERC advanced grant SEPtiM). This study was supported by LabEx UnivEarthS (ANR-10-LABX-0023) and Idex Université Paris Cité (ANR-18-IDEX-0001), and by IPGP multidisciplinary program PARI, and by Paris–IdF region SESAME Grant no. 12015908.

Contributions

F.M. designed the research, performed the experiments and analyses, analyzed the data and drafted the manuscript. A.S. conceived the research, contributed to data analysis and revised the manuscript. J.B. contributed to the experiments, data analysis and revised the manuscript. E.D.Y contributed to data analysis and revised the manuscript. A.S. contributed to the Raman analyses and edited the manuscript. J.W. performed the NanoSIMS analyses and edited the manuscript. S.B. operated the FIB at IPGP. S.V. operated the FIB at EPL. E.B. operated the EMPA at EPL. N.W. contributed to the experiments at IPGP.

Extended data

Table S1: Experimental conditions for the high temperature runs. The thermal pressure was calculated using a conversion factor of 2.7·10$^{-3}$ GPa/K, as obtained in [1] for a similar starting material. The number is for a pure silicate in a solid pressure medium, as the effect of the hydrogen in unknown we consider this the higher boundary for pressure and the before heating value the lower one.

| Sample | Pressure (GPa) before heating | With thermal pressure (GPa) | Temperature (K) ± 150 | Duration (s) |
|---|---|---|---|---|
| Exp_09 | 46.2 | 57.5 | 4500-4600 | 10-15 |
| Exp_10 | 60 | 71.9 | 4700 | 30* |
| Exp_11 | 16 | 26 | 4000 | 15-20 |
| Exp_12 | 22.4 | 32.7 | 4100 | 20 |
| Exp_13$^+$ | 30.7 | 41.23 | 4100-4200 | 20 |

* few seconds into heating a big flash is observed, then the sample heats homogeneously.
$^+$ the temperature measure software did not save the data. Reported numbers come from the logbook.

Table S2: Composition of the starting material and recovered experiments, available as a separate excel file. Volatile content reported as water, as analyzed with the NanoSIMS. As it was not possible to determine the speciation of hydrogen in the melt we also report the value as $H_2$ and $OH^-$.

Text S3: The boundary between the melt portion and the region with crystallized solids was recognizable in the samples at high magnification. The presence of nanometric bubbles in the melt and a higher Ca content also helped in confirming the identification. In laser heated diamond anvil cell, upon quenching, the central melt portion of the sample, does not quench to a homogeneous glass, but crystallizes in nm-sized crystals with chemistry corresponding to the stable mineral phases at the P-T conditions of the experiments[2]. The crystallized residual melt is not the exact structural and chemical representation of the melt at high temperature and processes leading to changes in the hydrogen content can and will take place during quenching. For instance, if the hydrogen is present as hydroxyl group it could be retained upon quenching. Conversely if it was present as molecular $H_2$ part it would exsolve and evaporate due to its incompatibility in lower mantle minerals. Therefore, the hydrogen concentration measured here is a lower bound on the real value in the melt at high temperature.

Text S4: The texture observed in the Fe-enriched blebs though it can be attributed to quench, as in other diamond anvil cell studies, has sufficient diverging features to question this attribution. In particular, all the visible bubbles have different size, are randomly distributed and in some cases (as for Exp_11) also contain volatiles along with the silicate (inferred from the presence of resin). The uncertainty about the process that led to the formation of these bubbles, hamper the possibility of using the chemical analyses on the blebs to determine the amount of oxides that were reduced. Because we can't tell if the observed composition is representative of the element content at the experimental conditions or not. If the value for each element reported in Figure S4 is indeed representative of the oxides' reduction, then along with iron MgO and $Al_2O_3$ would contribute 2 wt% water, and $SiO_2$ 12 wt% to the total water content.

Text S5: Raman analyses were performed only on the control sample. Data were collected from the top, with the gasket still on the bottom part of the cell, and after FIB milling, on the transverse section. In both geometries data were collected on an area and on multiple spots. Patterns are available as a separate excel file (Raman_data).

Figure S1: Pictures taken with the SEM before and during FIB milling, from the top and from the side, illustrating the different steps of samples preparation for nanoSIMS. The experimental chamber is located using the orientation marks. Its position is used to mill the first trenches, leaving a portion of sample with 30-25 µm thickness. Afterward the sample is milled alternating the two sides depending on which section is best suited for NanoSIMS analyses, milling proceeds until the sample reaches a thickness of 3-4 µm. The green and red boxes, as well as the green lines should be disregarded by the reader.

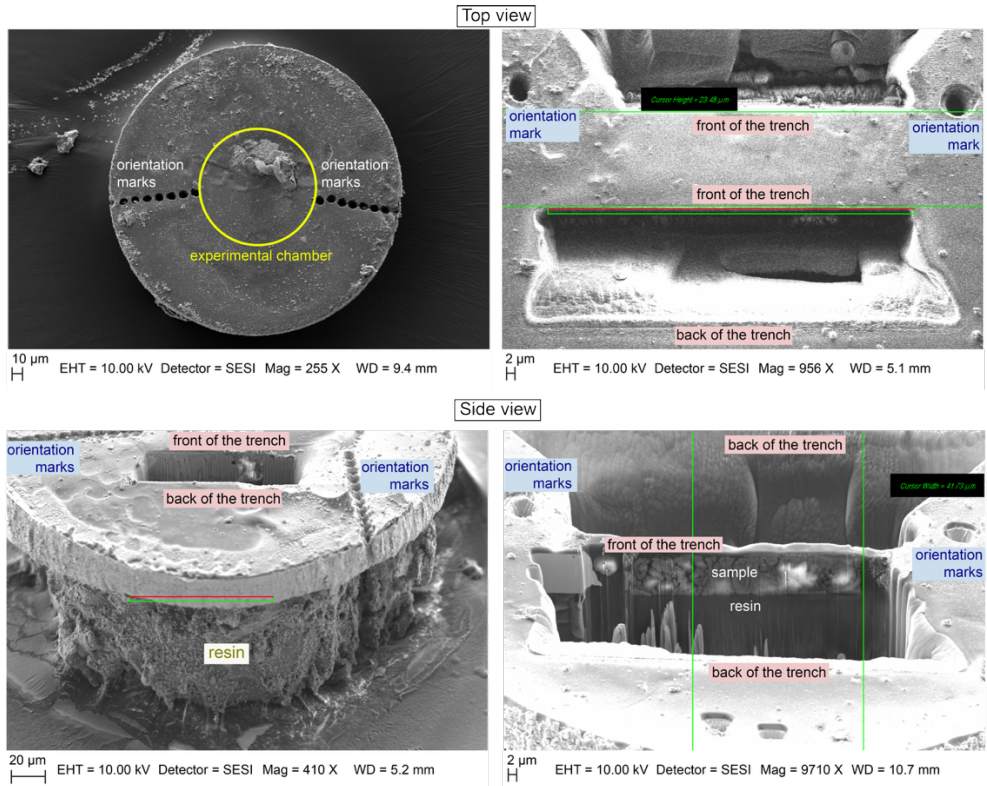

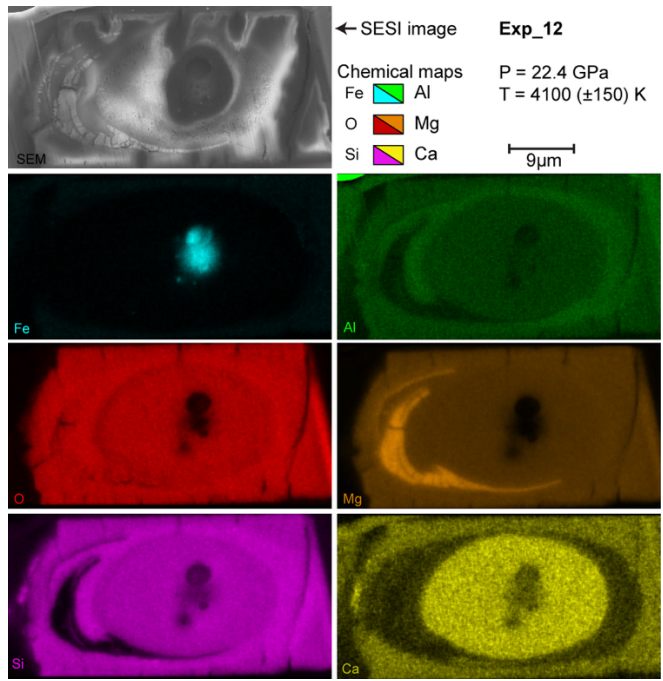
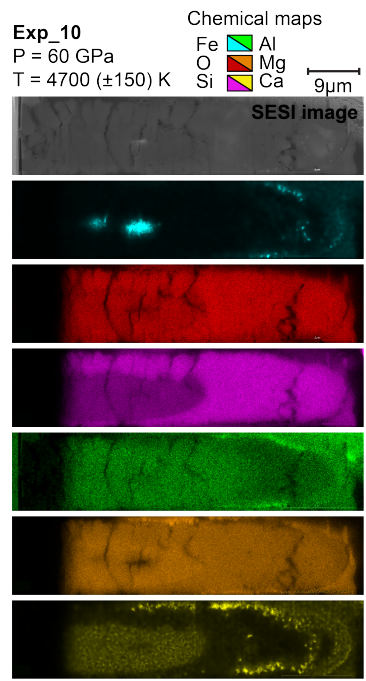

Figure S2: Composite image of chemical maps collected on two of the samples. Left: Example of a lower pressure sample. Right: example of a higher-pressure sample. The dark portion on the left-hand side is the shadowing imposed by the TEM grid. The concentrated Fe signal visible in the center of the melt is produced by the Fe-enriched blebs present on the other side of the thick section.

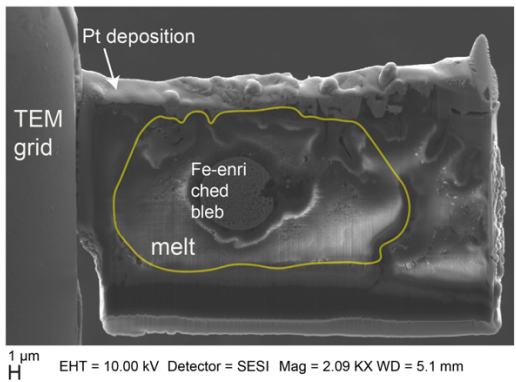

Figure S3: Sample Exp_13. From top, SEM image of the sample mounted on the TEM grid, SEM image of the sample and chemical map. In this sample, part of the FeO is still visible in the melt along with small scale (<1μm) Fe-enriched blebs. We infer that for this specific run the heating time was not sufficient for the reduction to consume all the FeO. The yellow line delimits the melt portion. Red circles are drawn around some of the small Fe-enriched blebs, some of which associated with empty bubbles (on the right side of the figure). Bubbles are visible also in the Fe-enriched bleb. Chemical analyses of the melt are reported in Text S2, chemical analyses of the metal are reported in Figure S4.

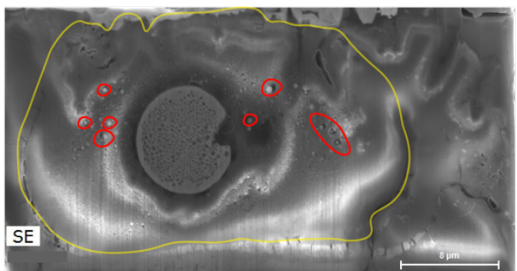
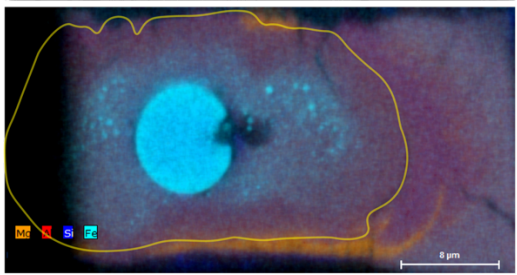

Figure S4: SEM images, magnification of the Fe-enriched blebs and chemical analyses for each area drawn in the enlarged images. Analyses are provided for the three samples showing the Fe-enriched blebs in the thick section chosen for NanoSIMS analyses. In Exp_11, the biggest bubble is observed to be filled by silicate material (light grey) and resin (black) which implies the previous presence of a volatile phase.

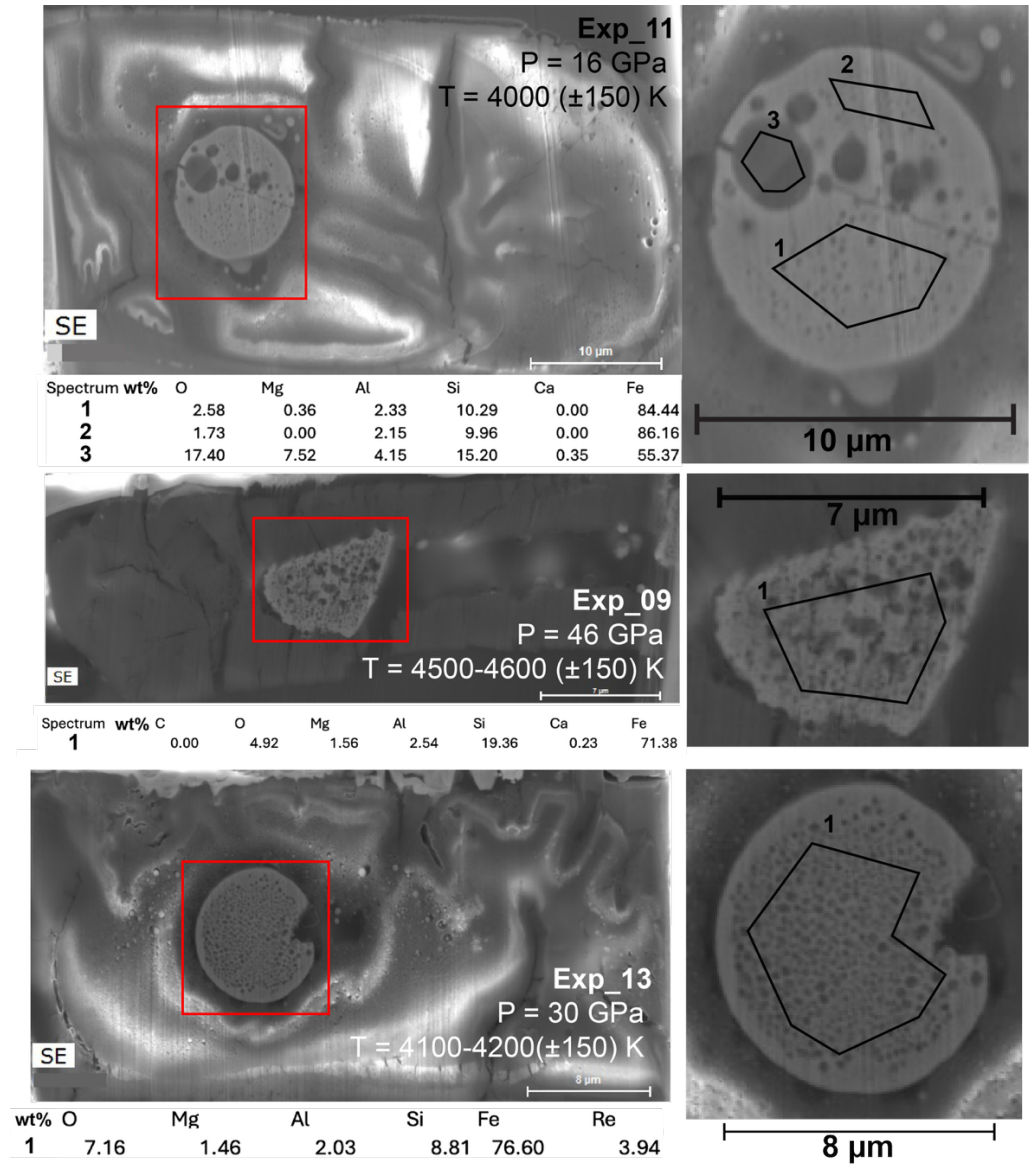

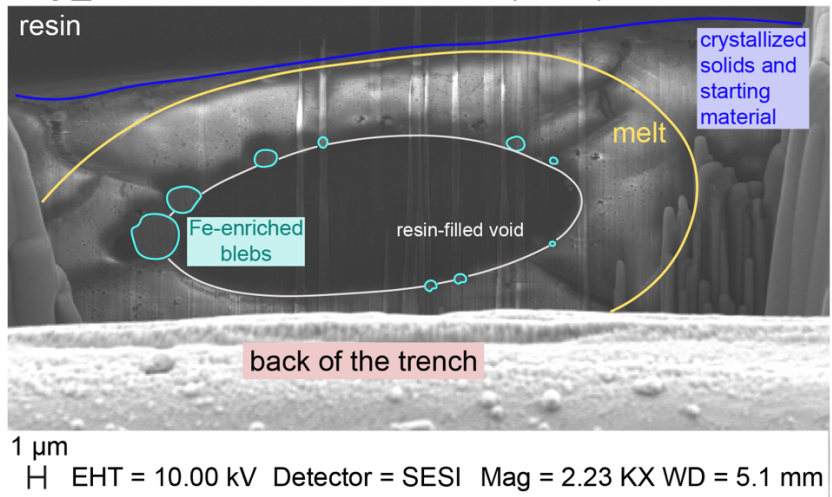
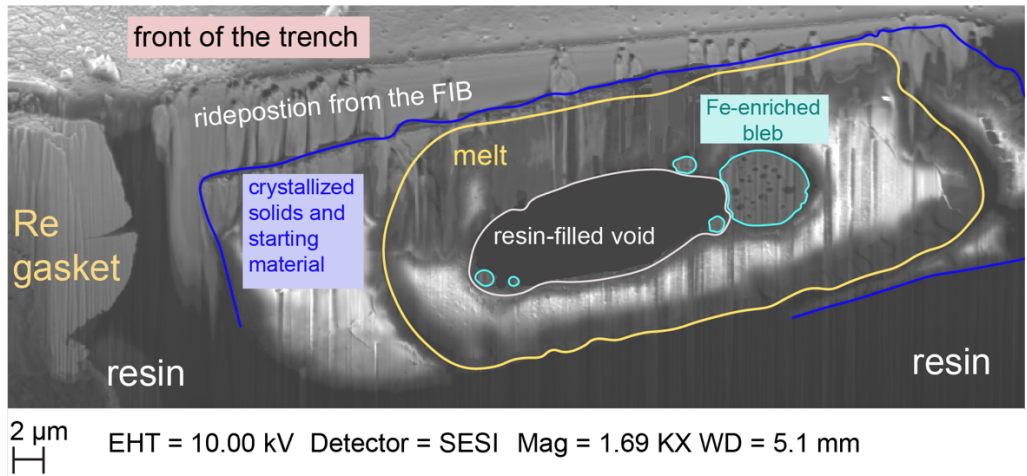
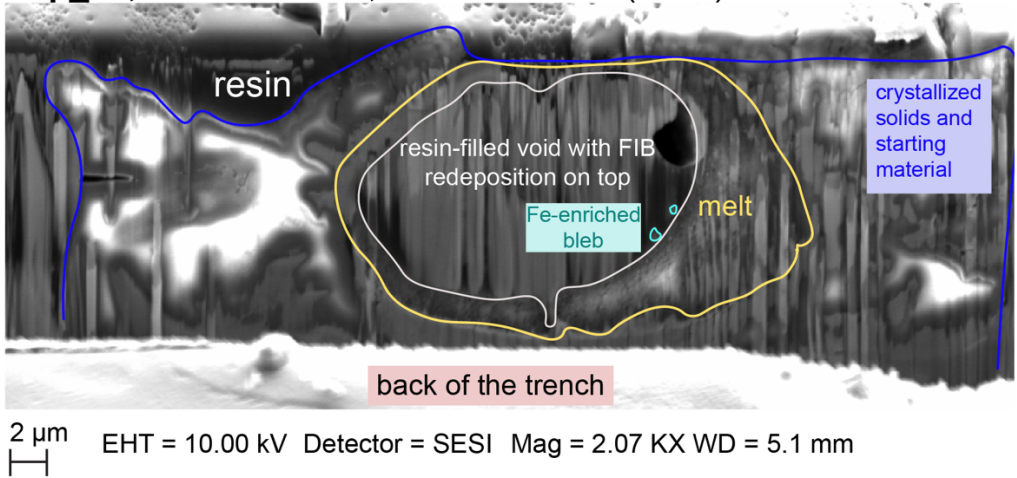

Figure S5: Images collected during FIB milling, on the lower pressure samples, when the cavities in the samples were at their maximum size. Colored lines serve as a guide for the eyes. Because the images were collected during the milling process some "artifacts" of FIB are visible. For examples the redeposition of milled material on the resin, visible in some cases (i.e., Exp_12 on top and the cavity of Exp_13) or the pillars visible on the bottom right side of sample Exp_11. In these lower pressure samples (i.e., Exp_11, 12, 13) the boundary between the Fe-enriched blebs, cavity and the silicate melt is well defined. The silicate melt looks homogeneous and consolidates with just few resin-filled fractures being visible.

See Figure S6 for the higher-pressure samples.

Figure S6: Images collected during FIB milling, on the two samples recovered from the experiments at the higher pressures. Exp_09, recovered from 46 GPa displays a long and elongated resin-filled cavity, irregularly shaped Fe-enriched blebs with fragmented borders and some fractures in the silicates. For Exp_10, recovered from the experiment at the highest-pressure images of two different sections, collected during milling, are provided. a) Was collected in proximity of the chosen section; b) was collected earlier in the process, when only a minimal portion of melt was visible. In this sample the much smaller, irregularly shaped and randomly distributed resin filled cavity are visible, along with the Fe-enriched bleb. The latter has an irregular shape and a silicate patch in the middle, suggesting the two phases are intertwined and dome-shaped. The silicate shows an extensive network of fractures and is poorly consolidated. In summary our observations suggest a progressive disappearance of the boundaries between the different phases, meaning the overpassing of the solvus and entering the pressure and temperature range of full miscibility between the present phases.

**Exp_09,** P = 46.2 GPa , T = 4500-4600 (±150) K

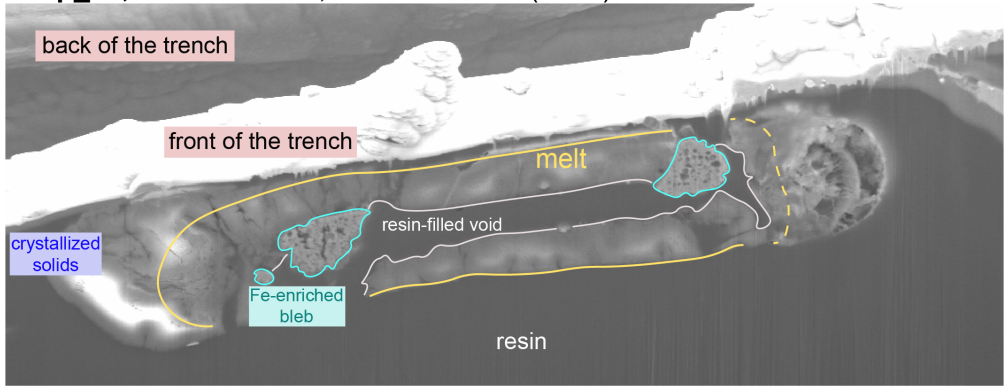

**Exp_10,** P = 60 GPa, T = 4700 (±150) K

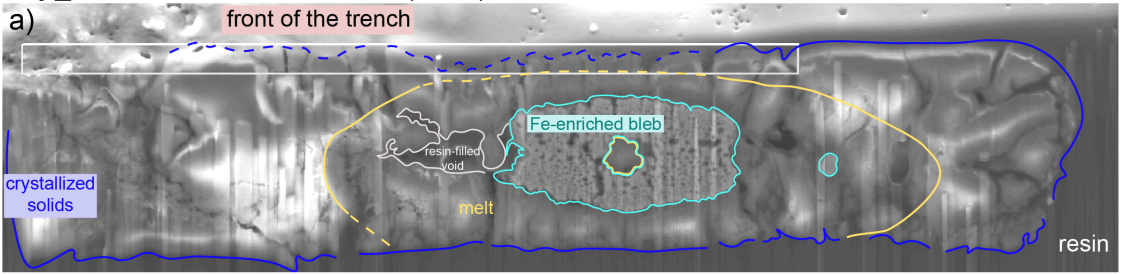

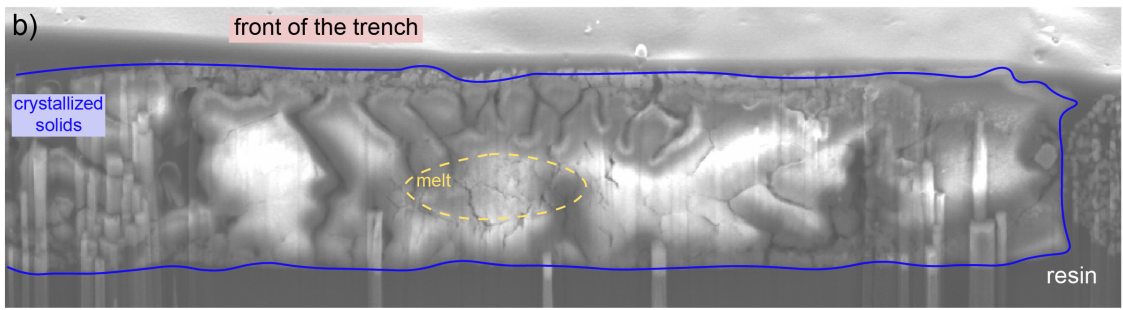

Figure S7: Data collected with the nanoSIMS reported as the ratio of the counts collected for channel 16OH and channel 30Si for all the samples. Examples of the ion maps used to identify suitable spots for the analyses are shown in Figure 1 and Figure S9.

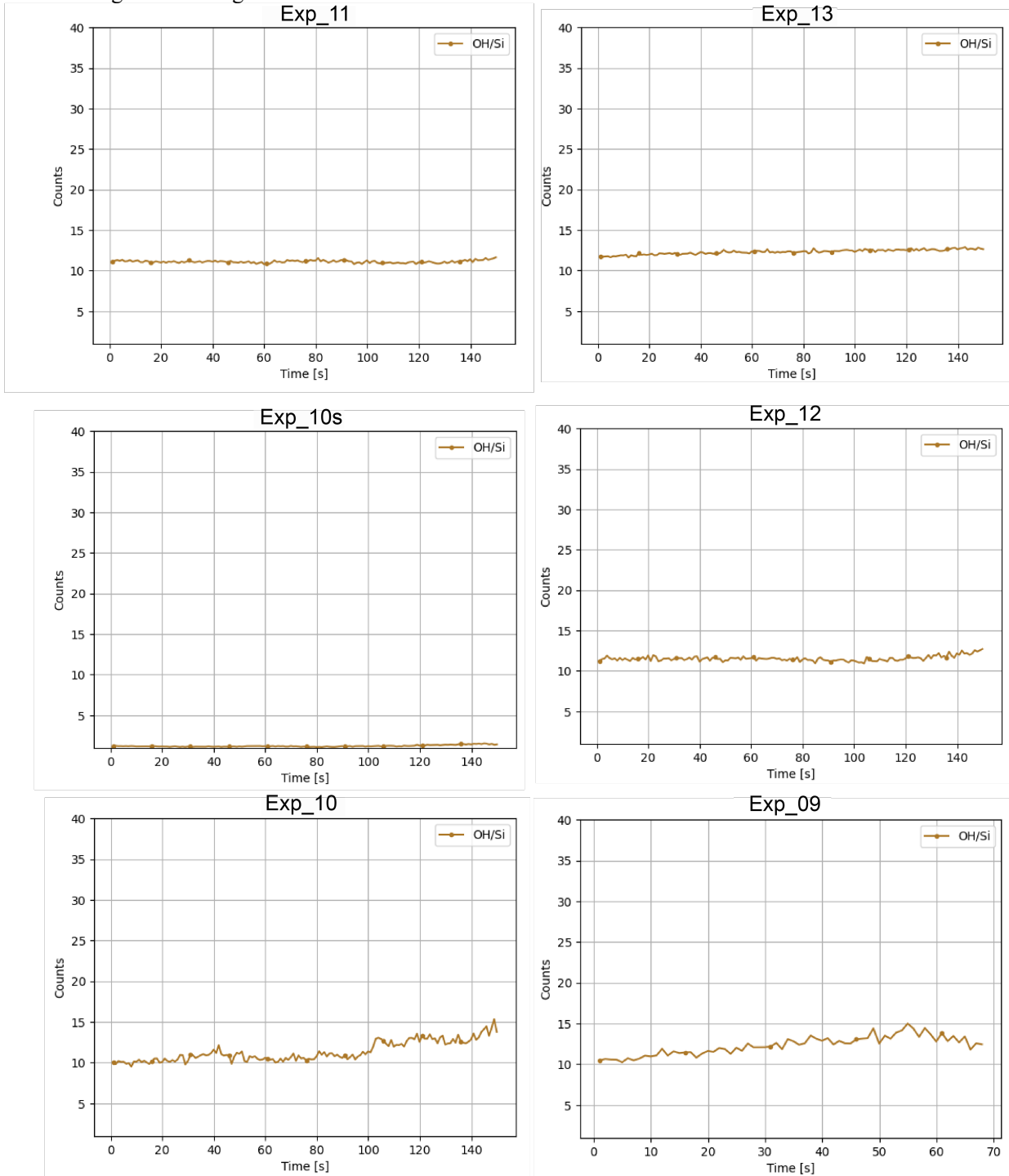

Figure S8: Experimental data shown as water content from the nanoSIMS as a function of pressure. Temperature is provided as the color scale.

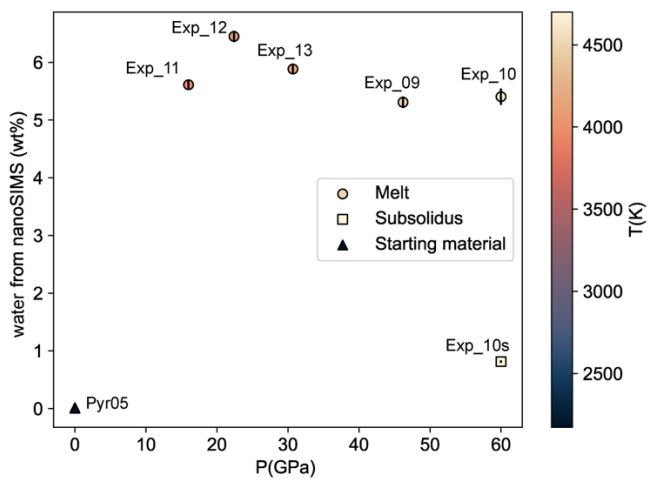

Figure S9: Left: Hydrogen map and SEM image on the same portion of the highest pressure sample. The box labeled "resin" on top is to show the portion where epoxy resin is present. Pink dashed line delimits the portion of the melt, yellow dashed line the crystallized solids. Right: Si map collected with the nanoSIMS along with the one collected with the SEM. A lower Si content is visible in the melt (inside pink dashed line) in both the images. NanoSIMS data on the melt were collected inside the pink line for the melt (Exp_10 in tables) and between pink and yellow lines for the crystallized solids (Exp_10s).

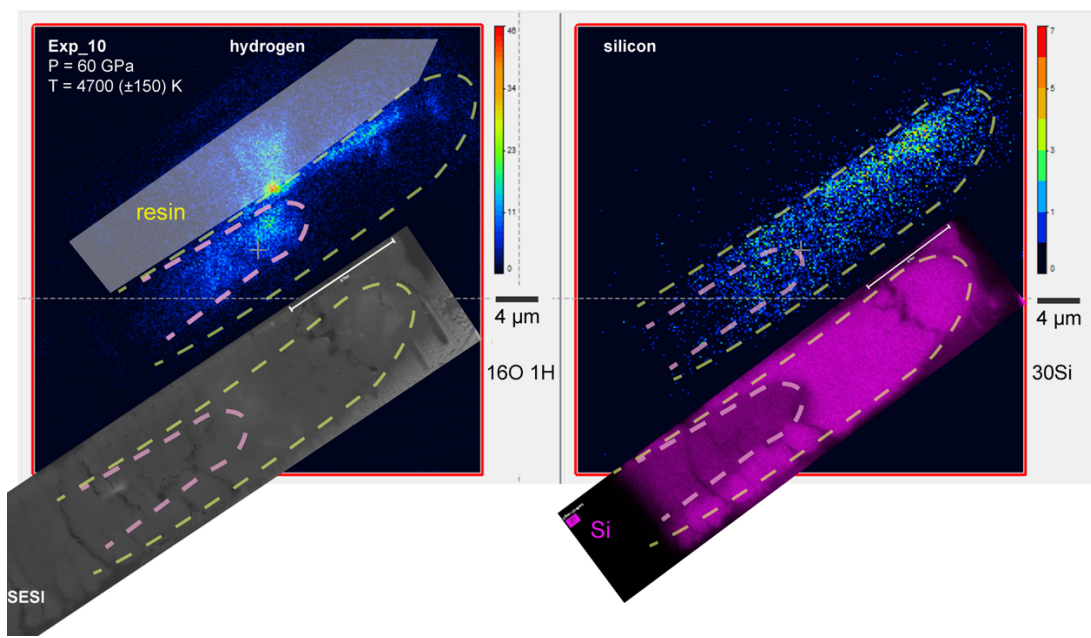

FigureS10: Comparison between the experimental data and those calculated with the model at the same pressure and temperature. The model reproduces the experimental data within the associated uncertainties.

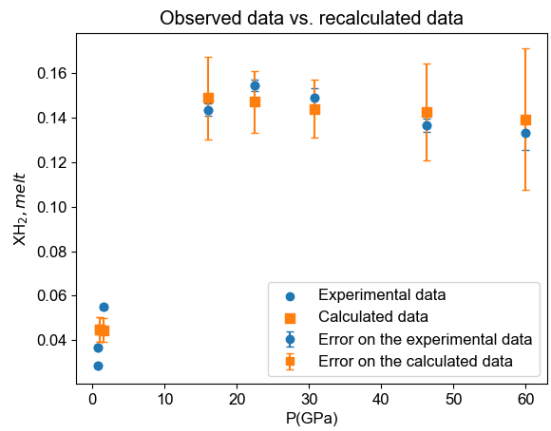

References:
1. Siebert, J., Badro, J., Antonangeli, D. & Ryerson, F. J. Metal – silicate partitioning of Ni and Co in a deep magma ocean. *Earth and Planetary Science Letters* **321–322**, 189–197 (2012).
2. Nabiei, F. *et al.* Investigating Magma Ocean Solidification on Earth Through Laser-Heated Diamond Anvil Cell Experiments. *Geophysical Research Letters* **48**, (2021).